\documentclass{mn2e}

\usepackage[dvips]{graphicx}
\usepackage{amssymb}

\begin{document}

\title[ Milky Way type galaxies]{Milky Way type galaxies 
in a $\Lambda$CDM
  cosmology}  
\author[De Rossi et al. ]{Maria E. De Rossi$^{1}$\thanks{E-mail:
  derossi@iafe.uba.ar}, Patricia B. Tissera$^{1}$, Gabriella De Lucia$^{2}$ and
  Guinevere \newauthor Kauffmann$^{2}$ \\ 
$^1$ Instituto de Astronom\'{\i}a 
y F\'{\i}sica del Espacio (Conicet-UBA), Argentina, CC67 Suc28, Buenos Aires
  (1428), Argentina.\\ 
$^2$ Max-Planck-Institut f\"ur Astrophysik, Karl-Schwarzschild-Str. 1, D-85748,
  Garching, Germany.\\ 
}
\maketitle
 
\begin{abstract}
  We analyse a sample of 52,000 Milky Way (MW)  type 
 galaxies drawn from the publicly available
 galaxy catalogue of the Millennium Simulation with the aim of studying statistically
the differences and similarities of their properties in comparison to our
Galaxy.
Model galaxies are chosen to lie in haloes with maximum circular velocities
 in the range 200-250 km s$^{-1}$ and to have bulge-to-disk ratios similar
 to that of the Milky Way. 
 We find that model MW galaxies  formed
 `quietly' through the accretion of cold gas and
  small satellite systems. Only $\approx 12$  per cent of our
 model galaxies experienced a major merger during their lifetime.
 Most of the stars
 formed `in situ', with only about 15 per cent of the final mass gathered
 through accretion. Supernovae and AGN feedback play an important role
 in the evolution of these systems. At high redshifts, when the potential wells
 of the MW progenitors are  shallower, 
 winds driven by supernovae explosions
 blow out a large fraction of the gas and metals.
 As the systems grow in mass, SN feedback effects decrease and
 AGN feedback takes over, playing  a more important role in   the regulation of the star formation activity at lower
redshifts.
 Although model Milky Way galaxies have been
 selected to lie in a narrow range of maximum
 circular velocities,  they nevertheless
 exhibit a significant dispersion in the  final stellar masses and  metallicities. 
 Our analysis suggests
 that this dispersion results from the different accretion
 histories of the parent dark matter haloes.
Statically, we also find evidences to support the Milky Way as a typical Sb/Sc
galaxy in the same mass range, providing a suitable benchmark to constrain
numerical models of galaxy formation.
\end{abstract}


\section{Introduction}
\label{intro}

Our own Galaxy (the Milky Way) has always represented a challenge for galaxy formation theories. 
Being the
only system for which we can access full phase-space information for a
significant number of individual stars, 
 the Milky Way
represents an important benchmark for theoretical models
(Perryman et al. 2001; Beers et al.
2004; Everdasson et al. 1993; Steinmetz et al. 2006; Ivezi\'c et al. 2008).
However, it is only one galaxy among many, 
and may not be representative
of `typical' spiral galaxies in the same mass range (Hammer et al. 2007).
Hence, a statistical analysis of MW type galaxies could help us to understand to which
extent we can rely on the Milky Way to set constrains for galaxy formation models.

The first model for the formation of our Galaxy was proposed by Eggen,
Lynden-Bell \& Sandage (1962), who argued that the observed relation between the
metallicity and the orbital eccentricity of a sample of about 200 stars could
be interpreted as a signature of a rapid radial collapse which led to the
formation of the stellar halo. Searle \& Zinn (1978) later proved this scenario
to be inconsistent with the observation of a negligible metallicity gradient
for the globular cluster population at large galactocentric distances. These
authors proposed an alternative scenario in which the stellar halo of the Galaxy
formed through accretion of smaller galactic systems.  This picture is in
qualitative agreement with expectations from the Cold Dark Matter (CDM) model
and with the observed signatures of substructure in the stellar halo of the
Milky Way, which appears to be a complex dynamical system still being shaped by
merging of smaller neighbouring galaxies (e.g. Vivas \& Zinn 2006).

Numerical simulations of structure formation in a CDM Universe
indicate the important role of the merging histories of dark matter haloes in
determining the structure and motions of stars within galaxies.
These simulations imply that the last major merger event in our
Galaxy should have occurred at $z >1$, otherwise the very thin cold disc
observed in the Galaxy would have been destroyed (Navarro et al. 2004; Kazantzidis et al. 2008). Bekki
\& Chiba (2001) have also shown that dissipative mergers with gas rich systems
could have generated halo stars before the formation of the Galactic disk.
More recent studies using an hybrid approach that combines $N$-body simulations
and semi-analytic techniques (Font et al. 2006; De Lucia \& Helmi 2008) have
suggested that the stellar halo of the Galaxy formed from the accretion of a
few relatively massive satellites ($10^{8} - 10^{9} M_{\odot}$) at early times
($>$ 9 Gyr). 

Although the basic cosmological paradigm appears to be well established, and
supported by a large number of observational results, our understanding of the
physics of galaxy formation is still far from complete. Within the currently
accepted paradigm, galaxies form when gas condenses at the centre of dark
matter haloes, which assemble in a `bottom-up' fashion with smaller systems
forming first and  merging later into larger structures. The evolution of the
baryonic components is dominated by complex physical processes (e.g. star
formation, supernovae and AGN feedback, chemical enrichment, etc.) which are
poorly understood from both the observational and the theoretical viewpoint.
The morphology, dynamics and chemistry of a galaxy is the result of many
intertwined processes.  In this complex framework, a number of questions still
remain to be answered: how did the Galaxy assemble? How `typical' is the
Galaxy in the Local Universe?  Which physical processes play a role in
determining its physical and chemical properties? Which kind of merging   
histories lead to the formation of galactic systems similar to our Galaxy?

In this work we will address some of these questions by taking advantage of one
of the largest cosmological simulations of structure formation carried out so
far, the Millennium Simulation, which is 
combined with a semi-analytic model of galaxy
formation (for a recent review on these techniques, see Baugh 2006). The aim of
this paper is to explore the formation histories of Milky Way-type galaxies and
to analyse the origin of the dispersion in their physical properties. In order
to achieve this goal we study simultaneously the assembly and chemical
evolution of model galaxies, and their location on the well-known correlation
between stellar mass and metallicity (e.g. Lequeux et al. 1979; Tremonti et al.
2004; Lee et al. 2006).  This strong correlation has been proved to evolve with
redshift in such a way that, at a given stellar mass, the gas-phase
metallicities of galaxies were lower 
in the past (e.g. Savaglio et al.  2005; Erb et al. 2006).  Studying the
evolution of the mass-metallicity relation as a function of cosmic time can
provide important information on the physical processes responsible for the
joint evolution of the chemical and dynamical properties of galaxies, e.g
supernovae and AGN feedback, star formation and mergers 
(e.g. Tissera et al. 2005; Brooks et al. 2007; De Rossi et al. 2007; Finlator et al. 2008).

This paper is organised as follows.  In Section 2 we give a brief description
of the simulation and of the semi-analytic model used in our study. In Section
3 we study the main physical properties of model Milky-Way type galaxies at
$z=0$, while in Section 4 we study their assembly and formation histories. In
Section 5 we analyse the influence of different assembly histories on the
chemical properties of model galaxies, as a function of redshift. Finally, we
summarise our findings in Section 6.

\section{The numerical simulation and the galaxy catalogue}
\label{sec:simsam}

This work takes advantage of the Millennium Simulation database\footnote{A
  description of the publicly available catalogues, and a link to the database
  can be found at the following webpage:
  http://www.mpa-garching.mpg.de/millennium/}. The Millennium Simulation
(Springel et al. 2005) follows $N=2160^3$ particles of mass $8.6 \times 10^{8}
M_{\odot} \, h^{-1}$ in a comoving periodic box of 500 Mpc $h^{-1}$ on a side, and
with a spatial resolution of 5 kpc $h^{-1}$ in the whole box. The cosmological
model is consistent with the first-year data from the Wilkinson Microwave
Anisotropy Probe (Spergel et al. 2003): ${\Omega}_{\rm m}=0.25$, ${\Omega}_{\rm
  b}=0.045$, $h=0.73$, ${\Omega}_{\rm {\Lambda}}=0.75$, $n=1$, and
${\sigma}_{8}=0.9$. Here, ${\Omega}_{\rm m}$, ${\Omega}_{\rm b}$ and
${\Omega}_{\rm {\Lambda}}$ denote the total matter density, the density of
baryons, and dark energy density at $z=0$, in units of the critical density for
closure (${\rho}_{\rm crit}=3 {H_{0}}^{2}/8 {\pi} G$). ${\sigma}_{8}$ is the
rms linear mass fluctuation within a sphere of radius 8 Mpc $h^{-1}$
extrapolated to the present epoch.

The simulation data were stored in 64 snapshots from $z=127$
to the present day. For each snapshot,
dark matter haloes were identified using a standard friends-of-friends (FOF)
algorithm with a linking length of 0.2 in units of the mean inter-particle
separation. The algorithm {\small SUBFIND} (Springel et al. 2001) was  used
to decompose each FOF group into a set of disjoint substructures. Only
substructures retaining at least 20 bound particles (i.e. corresponding to a
mass larger than $1.72 \times 10^{10} M_{\odot} \, h^{-1}$) were considered 
genuine and  used to construct merger history trees as
described in Springel et al. (2005) and De Lucia \& Blaizot (2007 - hereafter
DLB07). These merger trees represent the basic input to the semi-analytic
model that is used to generate the publicly 
available galaxy catalogues. 
This methodology was originally introduced by Springel et al.
(2001) and De Lucia, Kauffmann \& White (2004), and it has been recently
updated to include a model for the suppression of cooling flows by `radio-mode'
AGN feedback (Croton et al. 2006).  A more detailed description of the physical
modelling can be found in Croton et al.  (2006) and in DLB07.

As illustrated in Fig.~1 in De Lucia et al. (2004), each model galaxy 
is assumed to be made up of four different 
baryonic components: (i) {\it stars}; (ii) {\it
  cold gas}, which is available for star formation; (iii) {\it hot gas}, which
is available for cooling and only associated to galaxies sitting at the centre
of FOF haloes; and (iv) {\it ejected gas}, which is made up of material that is
temporarily ejected outside the galaxy's halo by supernovae winds. For the
interpretation of the results presented below, it is worth 
noting that the
model adopts an instantaneous recycling 
approximation (i.e. it neglects the delay
between star formation and the recycling of gas and metals from stellar winds
and supernovae), and that metals are exchanged between the different components
in proportion to the exchanged mass (for details, see De Lucia et al. 2004).
In addition, the model assumes that all metals produced by new stars are
instantaneously mixed with the available cold gas (i.e. the model assumes a 100
per cent mixing efficiency).

Finally, we remind the reader that the model used in this study follows 
dark matter substructures explicitly, i.e  the haloes within which galaxies
form are still followed after they are accreted onto 
larger systems. This scheme
leads to three different types of galaxies: 
Central galaxies of FOF groups are
referred to as `type 0', and are the only galaxies fed by
gas that is cooling radiatively
from the surrounding halo. Galaxies attached to distinct dark matter
substructures are called `type 1', and their orbits are 
followed by tracking the
parent dark matter substructure until tidal stripping reduces its mass below
the resolution of the simulation (De Lucia et al. 2004; Gao et al. 2004). The
galaxy at the centre of the dissolving substructure is not 
affected by tidal stripping, and is assigned a merging time
using the classical dynamical friction formula. These galaxies are referred to
as `type 2'.

\section{The sample of galaxies}

 In this work, we defined as Milky Way (MW) type galaxies those type 0 systems
which
inhabit haloes  
with  $200 <{\rm V}_{\rm max} < 250 {\rm \, km \, s^{-1}}$ (Li \& White 2008), 
where ${\rm V}_{\rm max}$ is the maximum circular velocity, which is estimated
directly from the simulation.  This selection provides
a total of 130311 systems, which is consistent with the number
of $L^*$ galaxies estimated from the observed local 
luminosity function. 
 We also restricted the sample to galaxies with $1.5
< \Delta M < 2.6$ with $\Delta M= M_{\rm bulge} - M_{\rm total}$ ($ M_{\rm
  bulge}$ and $M_{\rm total}$ are the bulge and total magnitude in the B-band),
so as to select galaxies with an Sb$/$Sc morphology (Simien \& de Vaucouleurs
1986). The final sample used in this study is made up of 52149 
systems that we called MW type galaxies.  We choose the maximum
 circular velocity and the morphological type to select MW haloes
because these two physical properties are  observationally better  determined
than the total stellar mass and cold gas fraction, for example.
We also 
extend the analysis including observational estimates on these two parameters and 
refer to this sample as a restricted MW sample (RMWS).
Adopting $f_g= 0.088 \pm 0.019 $ and $M* = (4.75 \pm 1.1) 10^{10} \,{\rm M}_{\odot} \, h^{-1}$
(Guesten \& Mezger 1982; Kulkarni \& Heiles 1987;
Boissier \& Prantzos 1999; Lineweaver 1999; De Lucia \& Helmi 2008), we find
that 10.7\% of the total model MW type galaxies would  constitute a closer representation of our
own Galaxy.

\begin{figure}
\begin{center}
\vspace*{0.5cm}\resizebox{6.5cm}{!}{\includegraphics{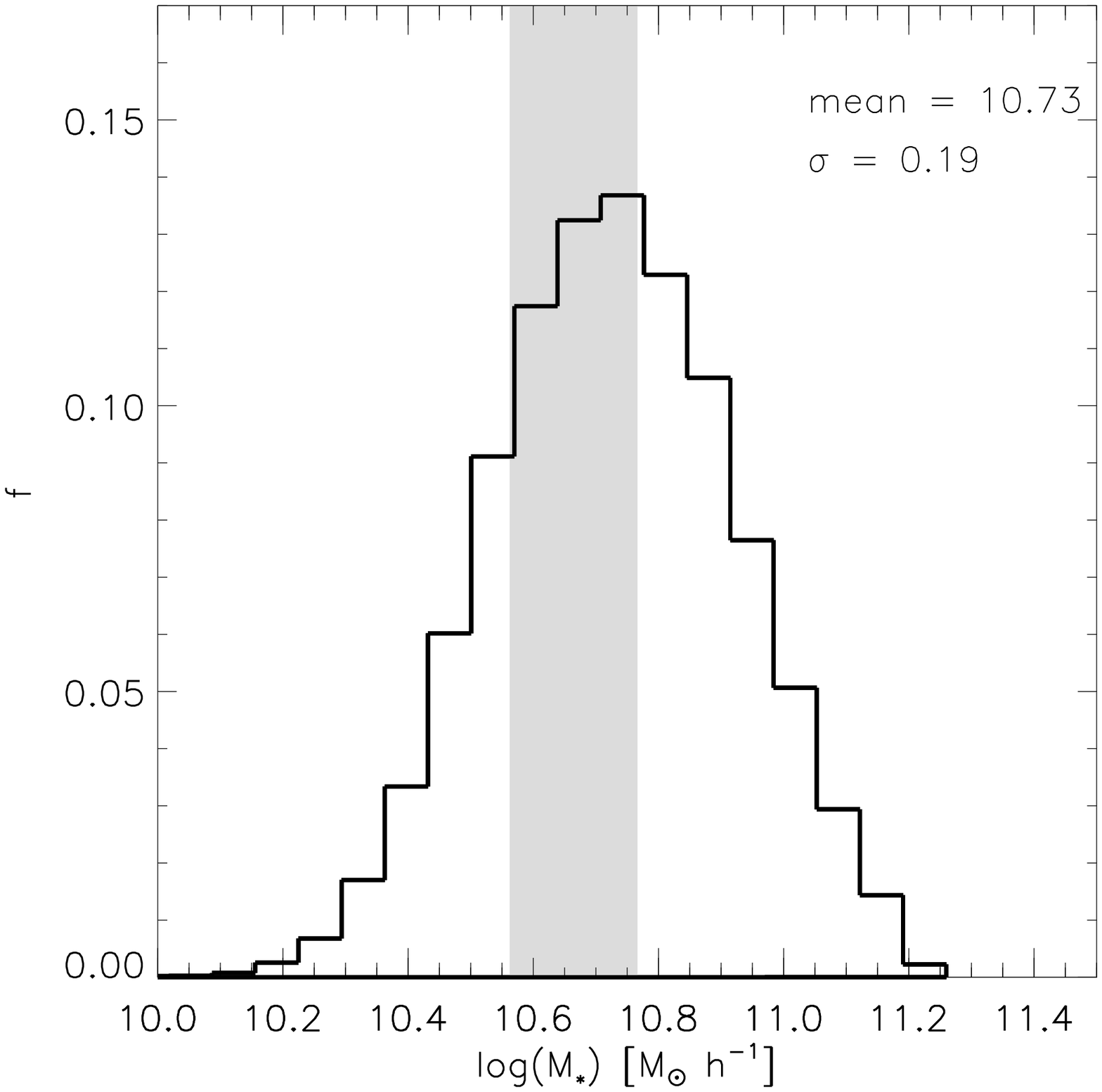}}\\
\vspace*{0.5cm}\resizebox{6.5cm}{!}{\includegraphics{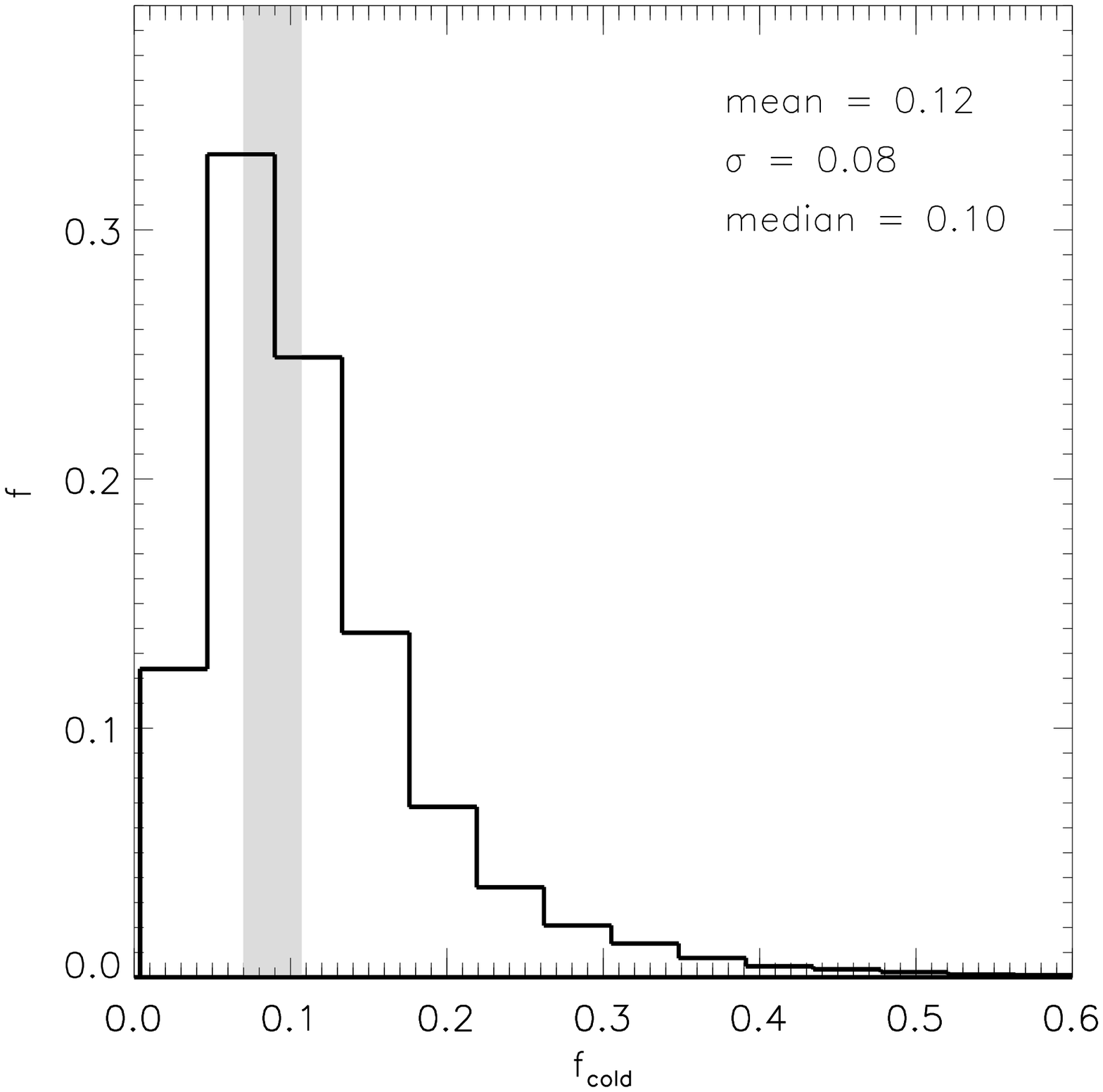}}\\
\vspace*{0.5cm}\resizebox{6.5cm}{!}{\includegraphics{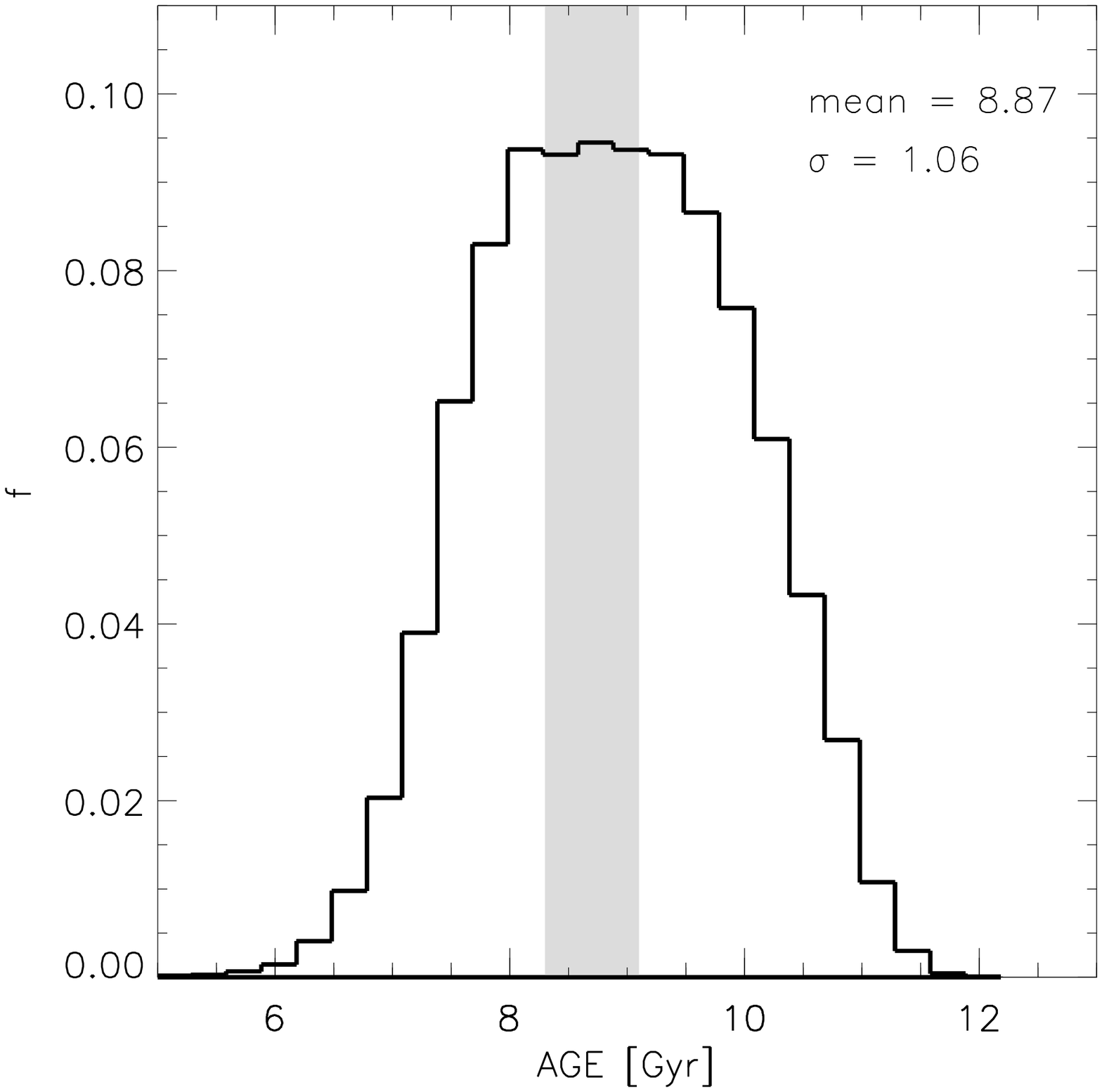}}\\
\end{center}
\caption{Distributions of stellar masses, cold gas fractions and stellar
  mass-weighted ages for the model MW type galaxies used in this study. The shaded
  regions correspond to the observational values for our Galaxy. See De Lucia
  \& Helmi 2008, and references therein. For the stellar age, an accurate
  total observational estimate is not available. For this quantity (bottom panel) we
  have plotted the range corresponding to the old thin disk by Lineweaver
  (1999).}
\label{histoDW}
\end{figure}

In Fig.~\ref{histoDW}, we show the distributions of total stellar masses ($M^*$),
cold gas fractions ($f_{\rm cold}$), and stellar mass-weighted ages for our model MW type galaxies.
The cold gas fraction is defined here as $\frac{M^{\rm cold}_{\rm gas}}{M^{\rm
    cold}_{\rm gas}+ M^*}$, where $M^{\rm cold}_{\rm gas}$ is the cold gas mass
of the system. 
The shaded regions in each panel show the available observational estimates for
our Galaxy (i.e. Guesten \& Mezger 1982; Kulkarni \& Heiles 1987;
Boissier \& Prantzos 1999; Lineweaver 1999; De Lucia \& Helmi 2008) 
\footnote {Note that while the mean ages of model
MW galaxies were calculated considering the whole stellar component, 
the observed values correspond only to the old thin disk of the Galaxy.}. 
The
distributions shown in Fig.~\ref{histoDW} peak close to these observational
estimates,  but also exhibit a large dispersion. This shows
that, although galaxies have been selected to have a narrow range of circular
velocities (i.e. total mass), they had significantly different evolutionary
histories, which produced the relatively large scatter in stellar mass and
gas content visible in Fig.~\ref{histoDW}.

\begin{figure}
\begin{center}
\vspace*{0.2cm}
\resizebox{8.5cm}{!}{\includegraphics{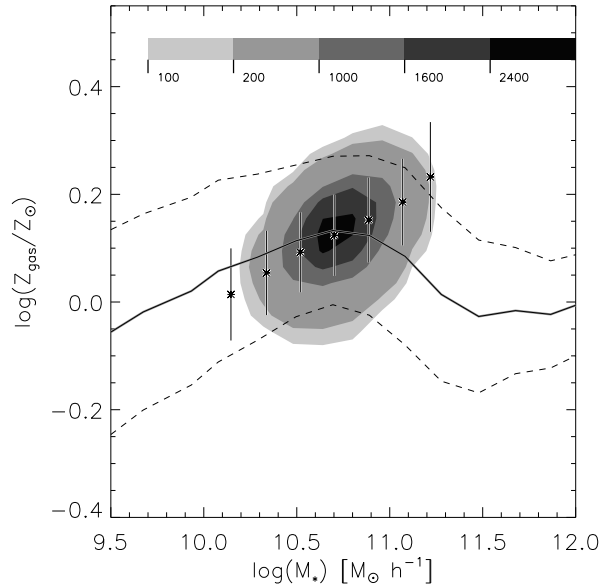}}
\hspace*{-0.05cm}
\end{center}
\caption{Cold gas-phase metallicity as a function of the stellar mass for the
  model MW type galaxies used in this study (contours), and for the full Millennium
  galaxy catalogue by DLB07 (solid line). The colour-coding of the contours
  indicate the mean number of galaxies in a bin of $0.15$~dex in stellar mass
  and $0.03$~dex in metallicity. Mean values and their standard deviations for
  the MW type galaxies are shown as symbols with error bars. Dashed lines indicate
  the dispersion in the Millennium MZR.}
\label{MZR_gral}
\end{figure}

In Fig.~\ref{MZR_gral}, we show the cold gas-phase metallicity as a function of
the stellar mass for the model MW type galaxies used in this study (contours).
Symbols with error bars indicate the mean and standard deviation of the
distribution. For comparison, we have also plotted the mass-metallicity
relation (MZR) for the full galaxy catalogue (the solid line indicate the mean,
and the dashed lines the scatter of the distribution). Fig.~\ref{MZR_gral}
shows that there is a well defined 
correlation between the gas-phase metallicity and the
stellar mass. The gas-phase metallicity increases approximately linearly for
stellar masses below $\sim 10^{11}\,{\rm M}_{\odot} \, h^{-1}$, and flattens for
more massive galaxies, remaining constant about the solar value. Model MW type
galaxies are located around the turnover of the Millennium MZR. Their mean
metallicity grows with stellar mass with a slightly steeper slope compared to
the general trend of the Millennium MZR, but with similar slope to that
reported by Tremonti et al. (2004) for the observed MZR of star forming
galaxies in the SDSS.  Model MW type galaxies show a standard deviation of $\sim
0.10$~dex, which is comparable to that estimated by Tremonti et al. We note
that the relatively large dispersion of the observed relation cannot be
entirely explained by observational uncertainties. In a recent work, 
 Cooper et al. (2008)  suggested that part (about 15 per cent) of
the scatter in the MZR could be explained by environmental differences. The
model relation shown in Fig.~\ref{MZR_gral} includes galaxies in all
environments (from field to clusters), and the
metallicities have  not been convolved with typical
observational uncertainties.

The MZR reflects the balance between star formation activity and other physical
processes such as Supernovae and AGN feedbacks, environmental effects,
etc. Studying the origin of this relation and of its dispersion can therefore
provide important information of the sequence of events which led to the
formation of galaxies similar to our own Milky Way. In the following, we will
analyse the origin of the MZR of model MW type galaxies by studying
their assembly and chemical enrichment histories simultaneously.

\section{The Assembly of model galaxies}

We constructed the  full galaxy merger trees for all our model MW type galaxies and
analysed them as in DLB07. We remind the reader that, among the different
`branches' of a galaxy merger tree,  the `main
branch' is special, because it connects a galaxy to its most massive progenitor
(the `main progenitor') at each node of the tree. In the following, we will
refer to all mergers onto the main branch as `accretion events'. Major mergers
correspond to accretion events in which the stellar mass of the accreted galaxy
is at least one third of the stellar mass in the main progenitor at the time of
accretion.

For the Millennium simulation, Springel et al. (2005) found that the halo mass function
can be reliably estimated down to 20 particles.
Moreover, De Lucia \& Helmi (2008) presented a complete study of numerical convergence for 
the semi-analytic model adopted in this work. These authors built up  galaxy merger trees and studied the
physical properties of the structure along them, 
in four realization of the same initial condition  but with increasing numerical resolution.
Their results indicate a very good level of convergence for the range of particle mass of few $10^8$  to $10^5 M_{\odot} h^{-1}$.
These previous studies provide evidences that for the mass range studied in this work, merger trees and the properties of galaxies along them are reliably
estimated.

\begin{figure*}
\begin{center}
\vspace*{0.5cm}
\resizebox{7.5cm}{!}{\includegraphics{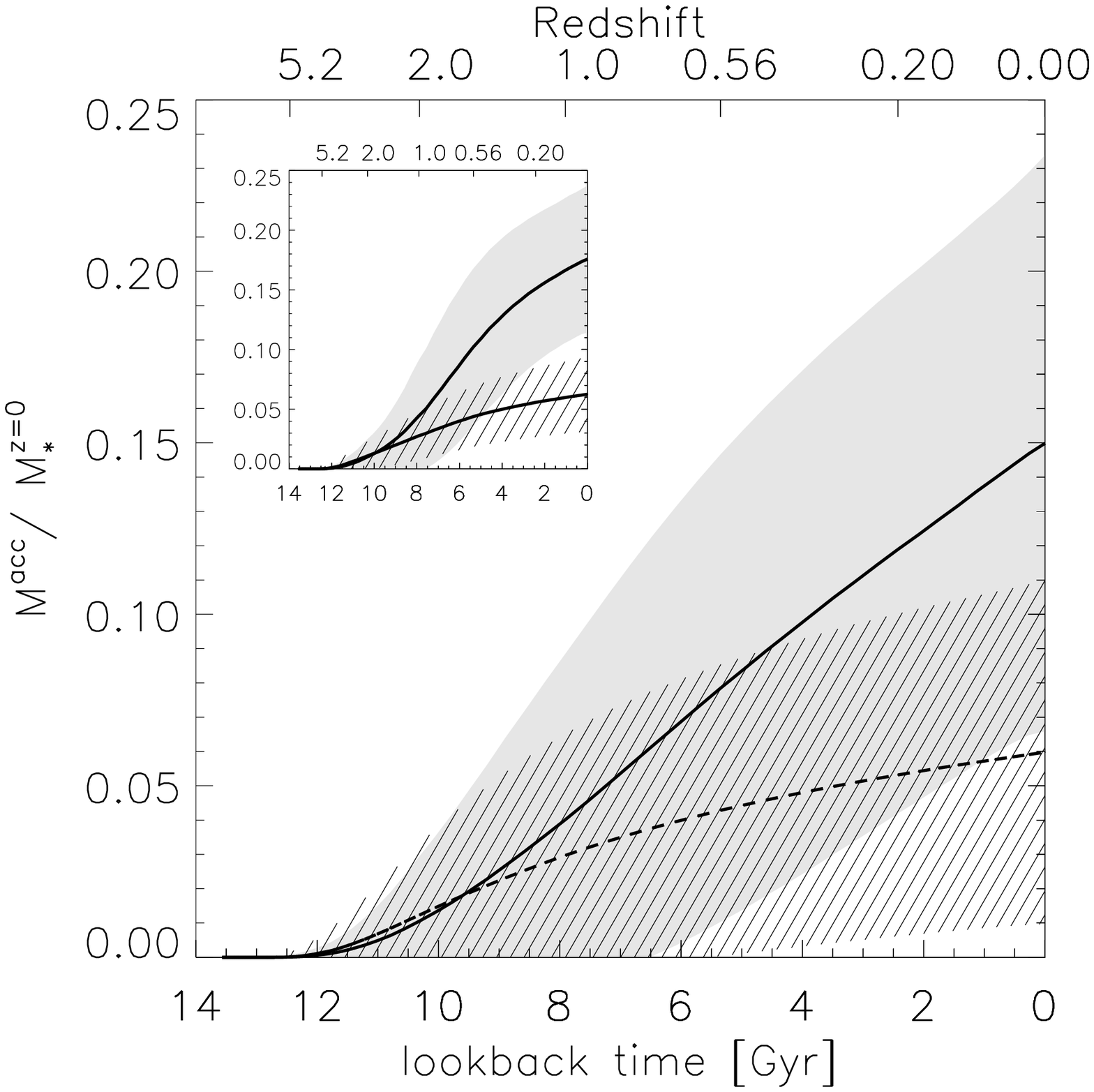}}
\hspace*{-0.2cm}
\vspace*{0.5cm}
\resizebox{7.5cm}{!}{\includegraphics{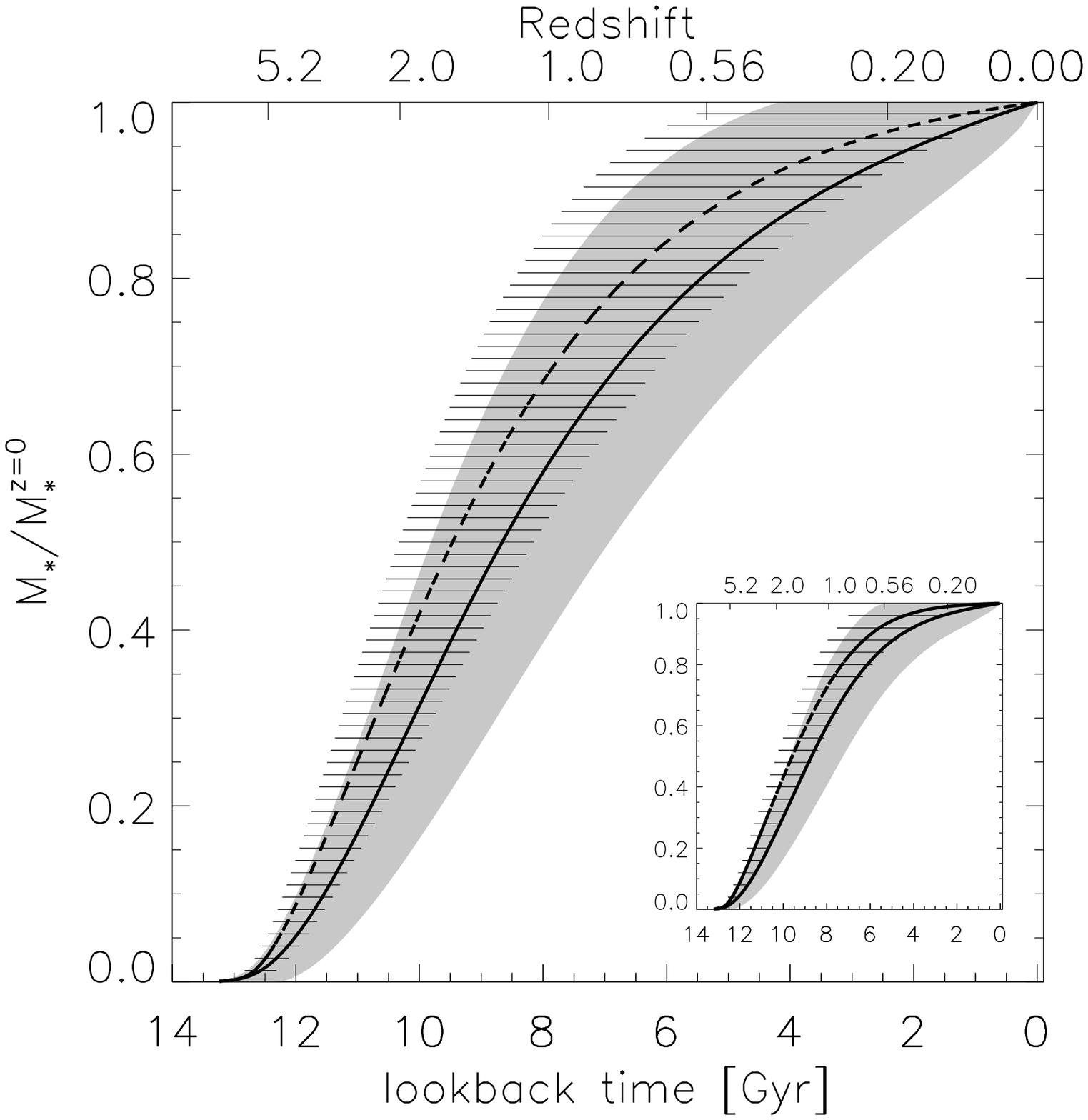}}
\hspace*{-0.2cm}\\
\vspace*{0.5cm}
\resizebox{7.5cm}{!}{\includegraphics{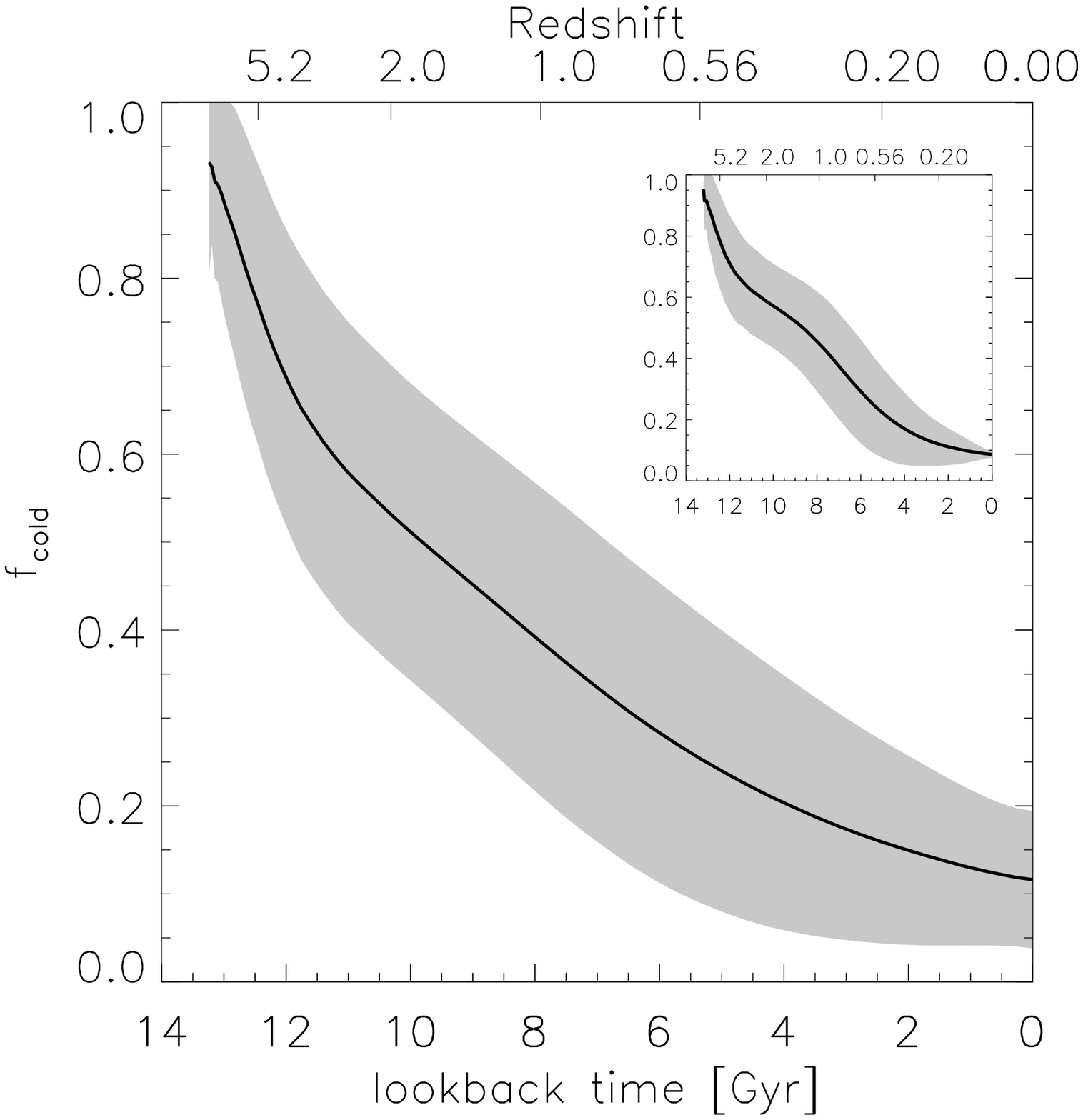}}
\hspace*{-0.2cm}
\vspace*{0.5cm}
\resizebox{7.5cm}{!}{\includegraphics{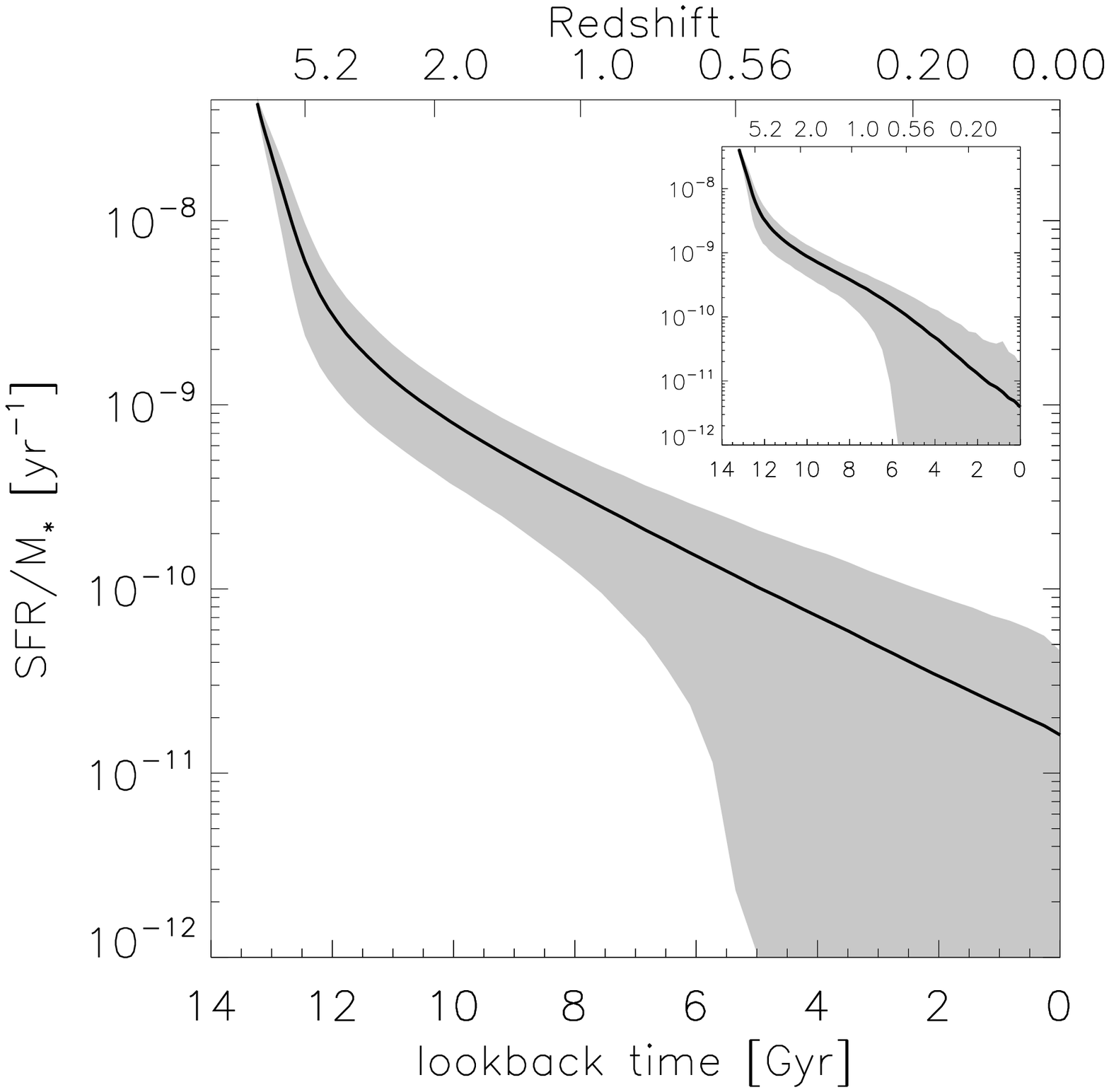}}
\hspace*{-0.2cm}\\
\end{center}
\caption{Top left panel: stellar mass (solid line) and cold gas mass (dashed
  line) accreted through mergers onto the main branch as a function of lookback
  time, normalised to the final stellar mass at $z=0$. Top right panel: mean
  evolution of the stellar mass in the main branch (solid line) and of the
  total stellar mass in all progenitors at any given time (dashed line).  Lower
  left panel: mean evolution of the cold gas fraction in the main branch.
  Lower right panel: mean evolution of the specific star formation rate in the
  main branch. In all panels, shaded (and dashed regions when present)
  indicates the 1-$\sigma$ dispersion of the distributions. The inset plots correspond
  to the results for the RMWS defined by imposing systems to fulfill constrains on  stellar masses and 
  cold gas fractions similar to our Galaxy.}
\label{MWmp_evol}
\end{figure*}

For our MW type sample, only a small fraction of the final stellar mass is
formed in the accreted systems. This can be seen in the top left panel of
Fig.~\ref{MWmp_evol} which shows the total stellar mass acquired through
mergers, normalised to the final stellar mass of the galaxy. At $z=0$,
the total accreted stellar mass varies between $\sim 7$ and $\sim 23$ per cent
of the final stellar mass, with a mean value of $\sim 15$ per cent. Most of the
accretion occurs relatively late, with only $\sim 5$ per cent of stellar mass
accreted before $z\sim 1$. The dashed region in the top left panel of
Fig.~\ref{MWmp_evol} shows the fraction of cold gas accreted through mergers as
a function of lookback time, again normalised to the final stellar mass. This
represents only $\sim 6$ per cent of the final stellar mass. The figure shows
then that about $80$ per cent of the stars in our MW type sample formed
{\it in situ}  in the main 
progenitors.  
Galaxies in the RWMS tend to 
have more important accretion events, particularly at z< 0.5, but differences 
with respect to the main sample are very small (less than 10\%).

The top right panel of Fig.~\ref{MWmp_evol} shows the mean evolution of the
stellar mass in the main progenitor, as a function of lookback time (solid
line), and the mean evolution of the total stellar mass already formed (i.e.
the sum of the stellar mass in all progenitors at a given time, dashed line).
The dashed and shaded regions show the scatter of the distributions. 
At $z \sim 1$, about 60
per cent of the total stellar mass is already in place in a single object, both 
for the main and the restricted sample.
Accounting for the stellar mass in all other progenitors does not
change these numbers significantly. 

The results shown in the top panels of Fig.~\ref{MWmp_evol} therefore
demonstrate that the evolution traced by following the main branch provides
a good representation of the evolution of our MW type galaxies. As discussed
above, most of the stars in the final systems are formed {\it in situ} through
the transformation of gas that comes primarily from infall. The mean evolution
of the cold gas component is shown in the bottom left panel of
Fig.~\ref{MWmp_evol}. The shaded region shows the scatter of the distribution.
The figure shows that mean gas fraction in the main branch declines from $\sim
0.8$ at $z\sim 5$ to $\sim 0.10$ at present. At $z=2$, the gas fraction is
$\sim 0.5$, in qualitative agreement with gas fractions measured by Erb et al.
(2006) for a sample of UV selected galaxies. At z=0, 
our typical MW type galaxy has a gas
fraction varying between $\sim 0.05$ and $0.20$, with a mean value of 0.12
(Fig.~\ref{histoDW}).
The trend is similar
for galaxies in the RMWS, but they exhibit a more rapid gas consumption between
$z \sim 1.5$ and $z \sim 0.5$.

The decrease of the available gas results in a decrease of the mean star
formation rate by about a factor of 10  between $z\sim2$
and $z\sim0.45$. The mean value of the star formation rate at present is about
$1.89 \,{\rm M}_{\odot}\,{\rm yr}^{-1}$, which is slightly higher than the 
$\lesssim 1\,{\rm M}_{\odot}\,{\rm yr}^{-1}$ observational
estimates for our Galaxy (e.g. Guesten \& Mezger 1982; Kulkarni \&
Heiles 1987; Boissier \& Prantzos 1999; Just \& Jahreiss 2007). 
The bottom right panel of
Fig.~\ref{MWmp_evol} shows the mean evolution of the specific star formation
rate. This exhibits a large spread, particularly at low redshifts,
 indicating
that a significant fraction of our MW galaxies have decreased
their star formation activity during the
last $\sim 6$~Gyrs. 
These effects are slightly stronger for galaxies in the RMWS (inset plot), which
end having  star formation rates
around $ \sim 0.26 \,{\rm M}_{\odot}\,{\rm yr}^{-1}$, which are now a bit lower
than the observational estimations  for our Galaxy.

This is (at least in part) due to the strong AGN feedback
adopted in the model in order to suppress cooling flows in relatively massive
haloes. This strong AGN feedback, combined with a relatively strong supernovae
feedback in this particular mass range, seems to exhaust the gas available  for star formation on relatively short
time scales (see also discussion in Sec. 6.1 of DLB07 for the brightest cluster
galaxies). A possible solution to this issue might be the implementation in the model 
of a new mechanism capable of re-activating cooling flows at later times but
this analysis is out of the scope of this work.

Only a small fraction (about 12 per cent) of galaxies in our MW type sample suffered
at least one major merger event during its life, i.e. all objects accreted onto
the main branch have  mass smaller than a third the mass of the 
main progenitor at the time of the merger event. A more detailed investigation
showed that most of the major mergers occurred at $z>1$, and
that only $3.23$ per cent of MW type galaxies had a major merger as their last
accretion event.  These results do not change significantly for galaxies in 
RMWS.
Therefore, our MW galaxies have a relatively quiet
life with no important mergers and a number of
minor accretion events that only add a relatively small fraction of the total
final mass.  

\begin{figure}
\begin{center}
\vspace*{0.5cm}\resizebox{8.5cm}{!}{\includegraphics{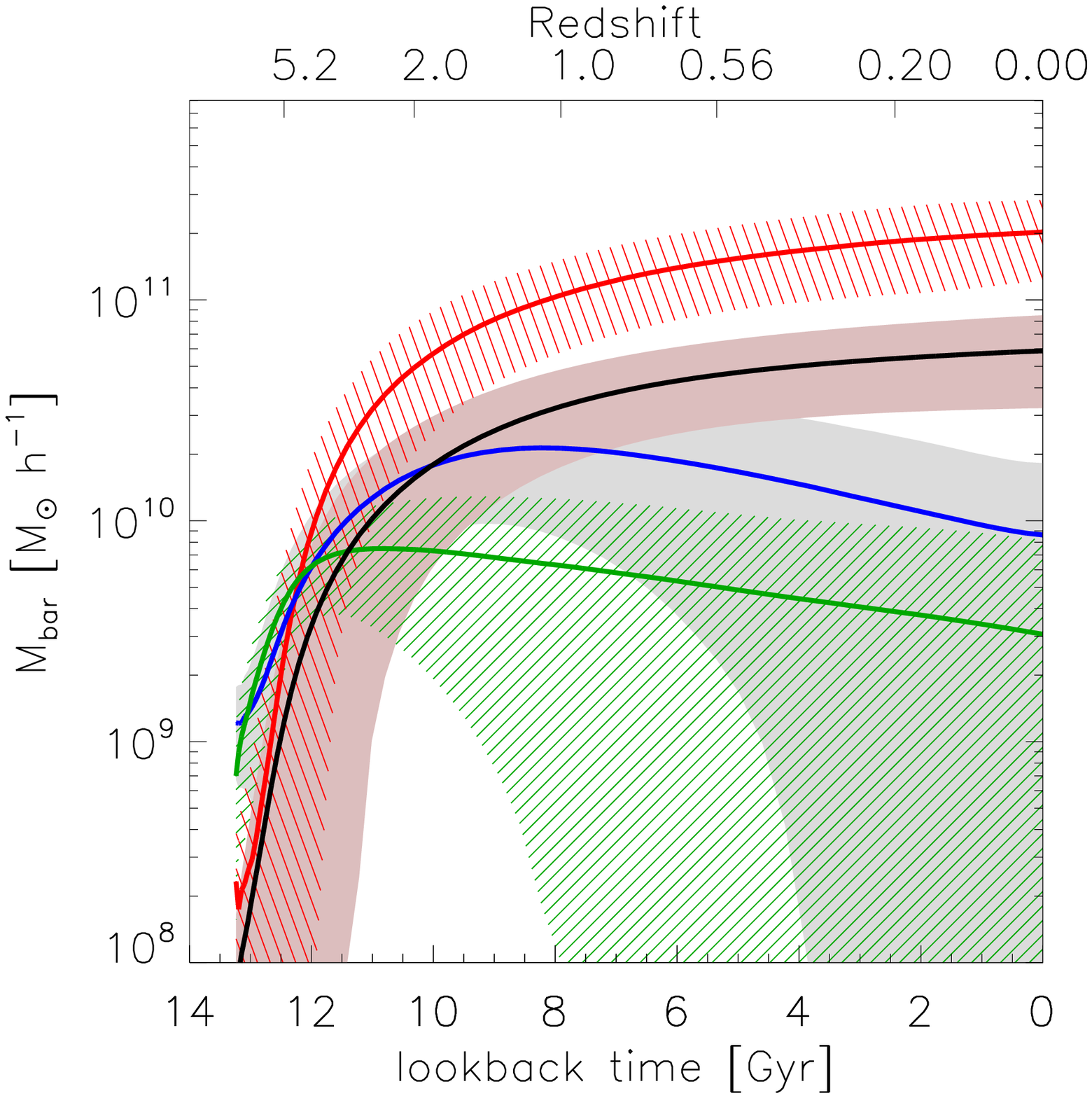}}\\
\end{center}
\caption{Evolution of baryonic mass for MW main progenitors.  
Note that the vertical axis has been plotted using logarithmic scale. The solid lines
  depict the mean relations for the different baryonic phases: cold gas (blue),
  ejected mass (green), hot gas (red) and stars (black).  The shaded and dashed areas
  represent the standard deviations.  
    }
 \label{Mbarvst}
\end{figure}

In Fig.~\ref{Mbarvst} we show the evolution of each of the baryonic phases (see
Sec.~\ref{sec:simsam}) associated to our MW galaxies as a function of
lookback time. 
Note that the vertical axis has been plotted using logarithmic scale.
It is clear that at lower redshifts most of the baryons are in the hot and
stellar phases, while the cold and ejected gas components dominate at very high
redshifts.  In particular, the ejected mass reaches a maximum at $z \approx 2.6$.
 In the
case of cold gas mass, it has a maximum at $z \sim 1.5$, when $f_{cold}$ is
around 0.4.  
Similar results are obtained for galaxies in the RMWS. 
In particular, the median cold gas mass at $z=0$ for the RMWS is approximately $6 \times
10^{9} M_{\odot}$, in good agreement with observations (Blitz 1997 and
references therein).
At $z<3$ we estimate quite large hot gas masses. This in part due to
the cooling flow suppression by AGN feedback used to built up the galaxy catalogue.
Croton et al. (2006) tuned the model to reproduce the local
relation between the mass of the bulge and the mass of the black hole 
 and it is known that the MW black hole is offset from this
relation.
Then, as expected, for our MW type galaxies we obtained a mean central black hole mass
of $\sim 4.8 \, \times \, 10^7 {\rm M}_{\odot} h^{-1}$, which is
one order of magnitude above the observational estimations for 
the Milky Way galaxy (Sch\"odel et al. 2002). 
Also  Supernova feedback contributes to build up the hot phase. 
Although at
lower redshifts supernovae are not the main source of heating, the excess of
hot gas might suggest that the combination of both type of feedbacks and their
relative importance need to be revised. Finally, we note that recent
observational studies have shown evidence for the existence of hot halo gas
surrounding spiral galaxies (Pedersen et al. 2006).

\section{Chemical evolution of model galaxies}

In the previous sections, we have studied the global properties of model MW
galaxies, focusing on their assembly and merger histories. In this section,
we will see how these histories influence the chemical evolution of MW
galaxies and their location on the mass-metallicity relation. 

\begin{figure*}
\begin{center}
\includegraphics{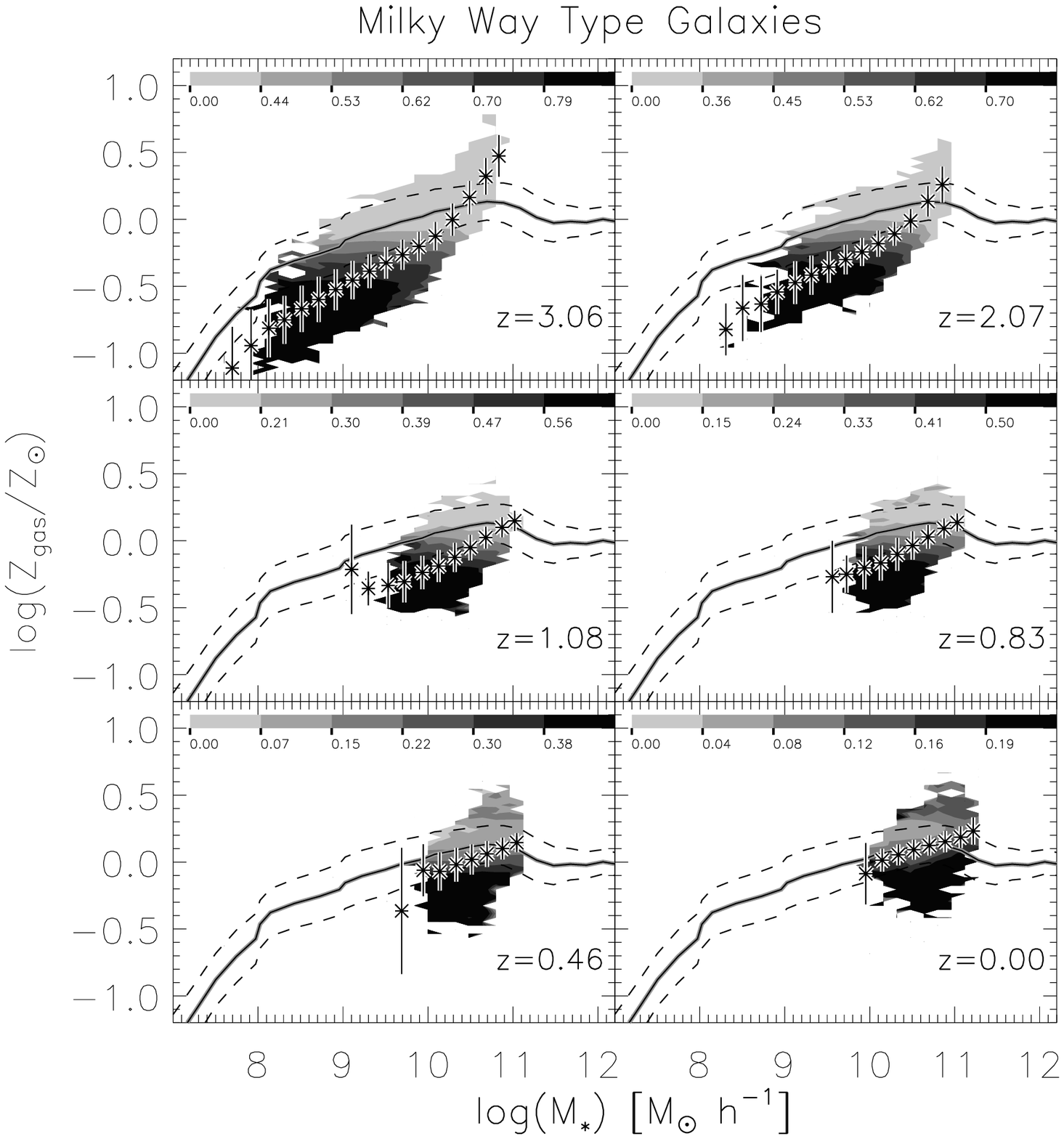}\hspace*{-0.05cm}\\
\end{center}
\caption{Evolution of the MZR for galaxies in our MW sample. At $z>0$, we plot
  the gas-phase metallicity and stellar mass of the main progenitor of each
  galaxy in the local sample. Contour levels are colour-coded according to the
  cold gas fraction. In each panel, we also show the mean (solid line) relation
  for all galaxies in the catalogue at $z=0$, and its dispersion (dashed
  lines).}
\label{2dfgc}
\end{figure*}

In Fig.~\ref{2dfgc} we show the MZR for the main progenitors of model MW
galaxies at different redshifts.  The distribution of
galaxies in this diagram is shown using contours , which have been
coded as a function of the cold gas fraction, $f_{\rm cold}$. Symbols
with error bars show the mean and dispersion of the distributions. As a
reference, we also plot the mean Millennium MZR (solid line) at $z=0$.

At high redshift, the MW progenitors cover a wider range of stellar masses
($\sim 3$~dex) and metallicities ($\sim 1.5$ dex) than the model MW galaxies at
$z=0$. Some progenitors reach gas-phase metallicities of around  solar or
even super-solar values at $z\sim 3$, while
others have significantly sub-solar values. 
 Fig.~\ref{2dfgc} also shows that most of the 
evolution in the metallicities of
the progenitors  occurs at $z>1$. Below this redshift, the distribution shrinks
significantly and the mean stellar mass of MW progenitors is typically larger
 than $60$ per cent of the final stellar mass of the systems (see also
Fig.~\ref{MWmp_evol} and relative discussion). 

As discussed in the previous section, the star formation rate in the main
progenitors of the model MW galaxies is tightly related to the availability of
cold gas as expected. Both of them regulates the chemical enrichment of the systems.
 At $z \approx 3$ our results   show that less massive (which are also  less
enriched) systems have, on average, larger cold gas fractions and
higher star formation rates.  
More massive systems have lower gas 
fractions and
star formation rates, which prevents further enrichment (Fig.~\ref{2dfgc}).
The largest changes in metallicity                               
occur for those systems which have the
lowest stellar masses (and consequently, the highest gas fractions) 
at high redshift. These galaxies exhibit an increase in
metallicity of $\sim 0.2$~dex since $z=1$. 
Galaxies in the RMWS exhibit the same general behavior, but 
covering a smaller parameter range
in the MZ-plane as expected.

\begin{figure*}
\begin{center}
\vspace*{0.5cm}
\resizebox{7.5cm}{!}{\includegraphics{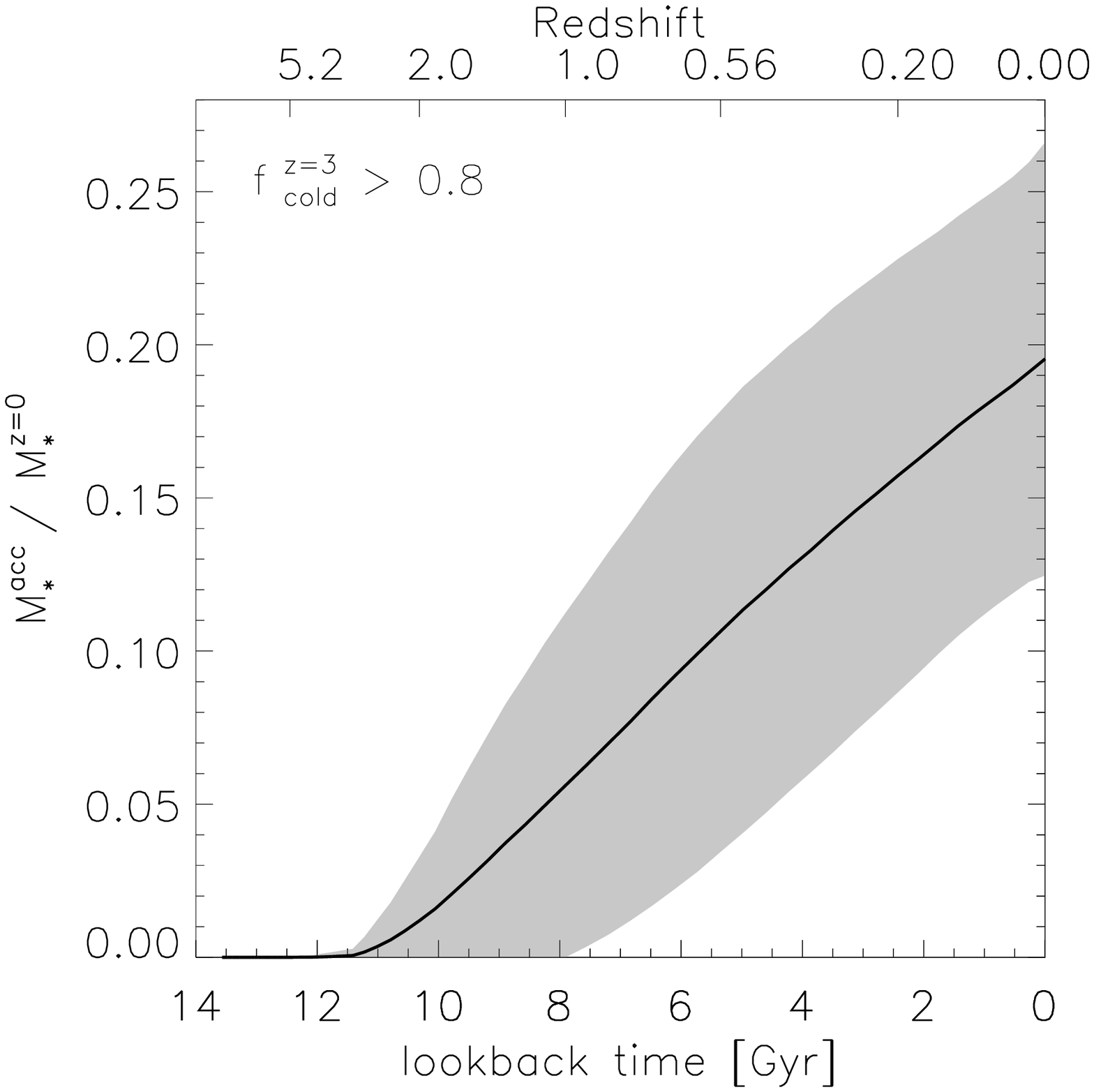}}
\hspace*{-0.2cm}
\vspace*{0.5cm}
\resizebox{7.5cm}{!}{\includegraphics{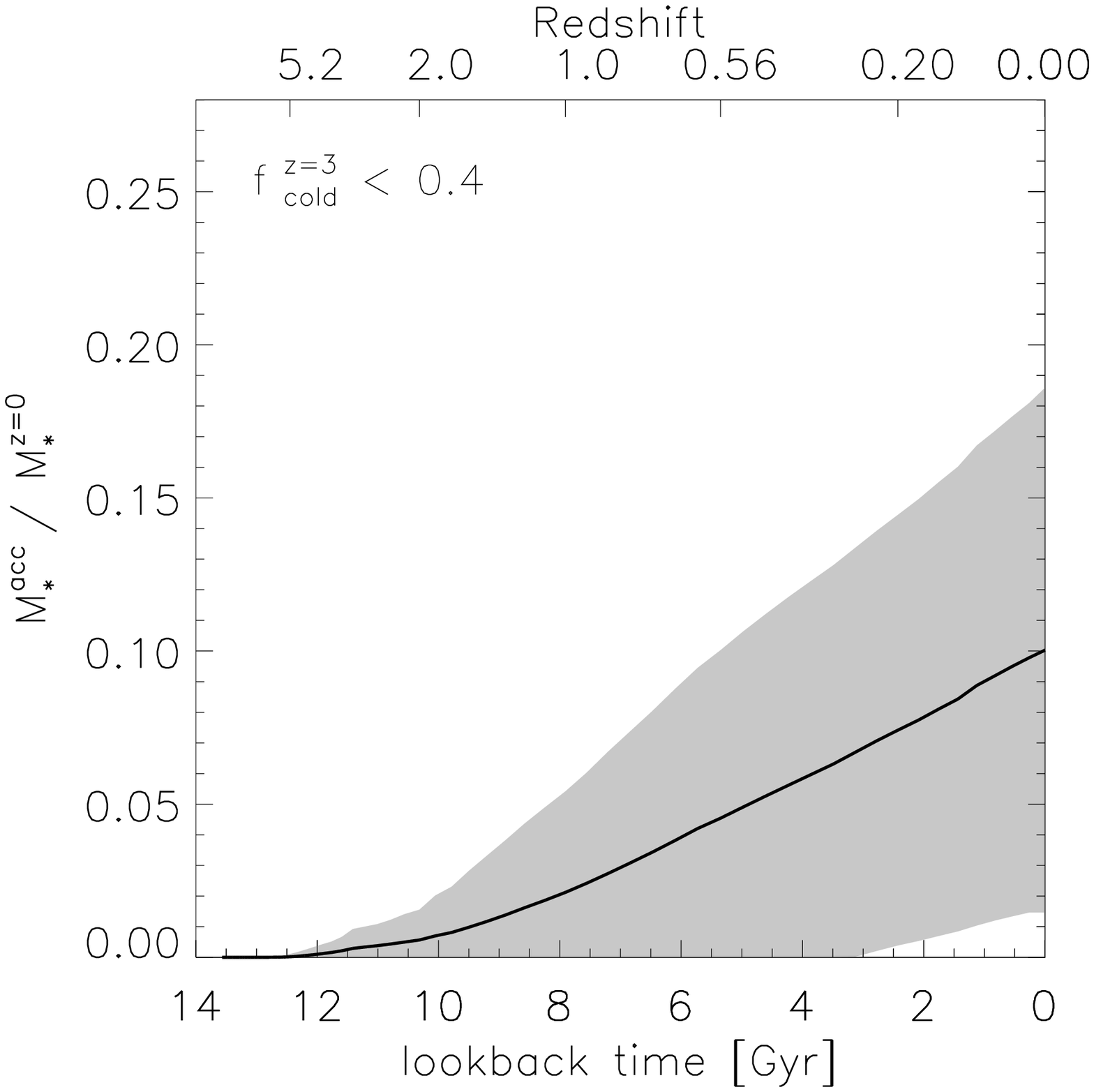}}
\hspace*{-0.2cm}\\
\end{center}
\caption{Evolution of the total accreted stellar mass for the main progenitors
  of MW type galaxies, normalised to final stellar mass at $z=0$. The left
  panel is for progenitors with cold gas fraction $>0.8$ at $z \sim 3$, while
  the right panel is for progenitors with cold gas fraction $<0.4$ at the same
  redshift.}
\label{fgc_z3_Ms}
\end{figure*}

In Fig.~\ref{fgc_z3_Ms} we show the mean fraction of mass accreted as a
function of lookback time for objects with large (left panel) and small (right
panel) gas fractions at $z\sim 3$. Gas-rich progenitors ($f_{\rm cold} > 0.8$)
not only have larger than average star formation rates, but also have
accreted a factor of two more stellar mass than gas-poor progenitors ($f_{\rm
  cold} < 0.4$), leading to more rapid chemical enrichment.  Conversely,
progenitors that are gas-poor at $z\sim 3$ do not evolve
significantly in metallicity;
they have completed most of their accretion already at
high redshifts (right panel of Fig.~\ref{fgc_z3_Ms}).

\begin{figure}
\begin{center}
\vspace*{0.5cm}
\resizebox{7.5cm}{!}{\includegraphics{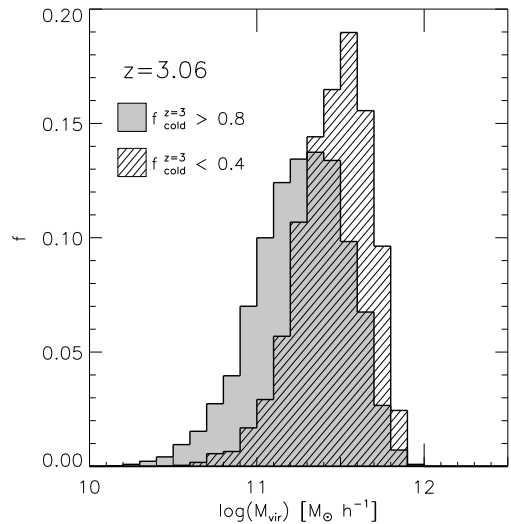}}
\hspace*{-0.2cm}
\end{center}
\caption{Distribution of virial masses of the main progenitors of MW type
  galaxies, at $z \sim 3$. Shaded histograms are for systems with cold gas
  fraction $>0.8$, while the dashed histograms are for systems with cold gas
  fraction $<0.4$ at the same redshift.}
\label{vir}
\end{figure}

The results discussed above indicate that the large dispersion in metallicity
and stellar mass of our MW galaxies is related to the  intrinsic dispersion
in the mass accretion histories of the haloes hosting MW type galaxies at $z=0$. Our
MW type galaxies were selected to lie in a  narrow range of  circular
velocities (i.e.total masses) and morphologies. Our results show that this
selection gives rise to a sample of galaxies with a relatively wide range of
stellar masses and metallicities, and that the
scatter can be related to the assembly and gas
accretion histories of the parent haloes. 
The halo mass thus does not uniquely     
determine the properties of the galaxies they host. This is
confirmed in Fig.~\ref{vir} which shows the distribution of virial masses of
haloes hosting the gas-rich and gas poor MW progenitors at $z\sim 3$. As we can
see, the two types of progenitors inhabit different mass haloes at high
redshift, although they end up in haloes of similar mass at the present day. 

\section{Conclusions}

In this paper we have studied the assembly and chemical evolution of MW type
galaxies. We used the galaxy catalogue built by DLB07 for the Millennium
Simulation, and selected MW type galaxies by imposing circular velocity and
bulge-to-disk ratio constrains.  Our model MW type galaxies have mean properties such 
 stellar mass, gas
fraction,  age and star formation rate,  in
good agreement with the estimations obtained  for our own Galaxy.
Interestingly, there is also substantial dispersion in these quantities which,
 accordingly to our work, can be related to their history of assembly.           

Taking advantage of the publicly available database
of merger trees, we studied the assembly
and chemical enrichment histories of our model MW type galaxies. 
Our main results can be
summarised as follows:

1. Most of the final stellar mass of model MW type galaxies is formed {\it in situ}
from gas infalling from the surrounding halo. A small fraction of the final
stellar mass (about 15 per cent) is formed in smaller galaxies
that are accreted over the lifetime of the Milky Way. 
Only 12 per cent of our model MW type galaxies experienced a
major merger during their lifetime, 
and for only 3.23 per cent of our sample was this major merger the 
last accretion event. 

2. The distribution of baryons in different components is regulated by feedback
processes. At $z > 4$, supernovae feedback is effective in ejecting large
fractions of the gas outside the haloes (due to the shallower potential
wells). By $z\sim 1$, a large part of this gas has been re-incorporated and the
suppression of cooling flows by AGN feedback starts playing a more important
role, keeping an important  fraction of the baryons in the hot phase. 

3. The MZR of model MW  type galaxies has a dispersion of $\sim 0.10$ dex, in
agreement with the observed results by Tremonti et al. (2004). We found that, at a given stellar mass,
the main parameter determining this dispersion is the
gas richness of the systems. Gas-rich systems tend to be more metal-poor, while
gas-poor galaxies have converted most of their cold gas component into stars in the past and, therefore
have  reached a higher level of chemical enrichment.

4. The accretion histories of the haloes hosting our MW type galaxies exhibit a
large dispersion.  Haloes
hosting gas-rich MW progenitors at high redshifts tend to experience a higher 
accretion rate  at later times. 
These differences in the accretion histories of the
parent haloes introduce differences in the  star formation rates
of the progenitors which, on their turn,   modulate the impact of
supernovae and AGN feedbacks. This leads to an important dispersion in the
stellar masses and metallicities of the z=0 Milky Way systems.

5.  If we restrict the model MW galaxies to satisfy also observational
constrains on stellar mass and gas fractions, we get a smaller sample
 with similar trends to those of the complete MW type sample reflecting the
fact that the dark matter halo is the dominant factor determining the history of
assembly of galaxies.  However, the dispersion in the metallicity and gas fraction
at a given mass is produced by slightly differences in the accretion rates of
substructures which regulate the star formation activity.

 Our findings suggest that  part of
the dispersion observed in the MZR could be revealing differences in the  histories of
formation. Also, our results suggest that the Galaxy may be consider a typical Sb/Sc
galaxy in the same mass range, providing  a suitable benchmark for numerical
models of galaxy formation.

\section*{Acknowledgements}
We thank the anonymous referee for her/his useful comments that largely helped to improve this
paper.
This work was partially supported by the European Union's ALFA-II programme,
through LENAC, the Latin American European Network for Astrophysics and
Cosmology.  We acknowledge support from Consejo Nacional de Investigaciones
Cient\'{\i}ficas y T\'ecnicas and Agencia de Promoci\'on de
Ciencia y Tecnolog\'{\i}a.  The Millennium Simulation databases used in
this paper and the web application providing online access to them were
constructed as part of the activities of the German Astrophysical Virtual
Observatory.


\begin{thebibliography}{}
\expandafter\ifx\csname natexlab\endcsname\relax\def\natexlab#1{#1}\fi
\expandafter\ifx\csname natexlab\endcsname\relax\def\natexlab#1{#1}\fi

\bibitem[\protect\citeauthoryear{Baugh}{2006}]{2006RPPh...69.3101B} Baugh 
C.~M., 2006, RPPh, 69, 3101

\bibitem[\protect\citeauthoryear{Beers et al.}{2004}]{2004IAUS..220..195B}
Beers T.~C., et al., 2004, IAUS, 220, 195

\bibitem[\protect\citeauthoryear{Bekki
\& Chiba}{2001}]{2001ApJ...558..666B} Bekki K., Chiba M., 2001, ApJ, 558, 666

\bibitem[\protect\citeauthoryear{Benson
et al.}{2000}]{2000MNRAS.314..557B} Benson A.~J., Bower R.~G., 
Frenk C.~S., White S.~D.~M., 2000, MNRAS, 314, 557  

\bibitem[\protect\citeauthoryear{Blitz}{1997}]{1997IAUS..170...11B} Blitz
L., 1997, IAUS, 170, 11

\bibitem[\protect\citeauthoryear{Boissier
\& Prantzos}{1999}]{1999MNRAS.307..857B} Boissier S., Prantzos N., 1999, MNRAS, 307, 857

\bibitem[\protect\citeauthoryear{Brooks et al.}{2007}]{2007ApJ...655L..17B} 
Brooks A.~M., Governato F., Booth C.~M., Willman B., Gardner J.~P., Wadsley 
J., Stinson G., Quinn T., 2007, ApJ, 655, L17

\bibitem[\protect\citeauthoryear{Cooper et al.}{2008}]{2008MNRAS.390..245C} 
Cooper M.~C., Tremonti C.~A., Newman J.~A., Zabludoff A.~I., 2008, MNRAS, 
390, 245 

\bibitem[\protect\citeauthoryear{Croton et al.}{2006}]{2006MNRAS.365...11C} Croton
D.~J., et al., 2006, MNRAS, 365, 11

\bibitem[\protect\citeauthoryear{De Lucia, Kauffmann,
\& White}{2004}]{2004MNRAS.349.1101D} De Lucia G., Kauffmann G., White S.~D.~M., 2004, MNRAS, 349, 1101

\bibitem[\protect\citeauthoryear{De Lucia et al.}{2006}]
{2006MNRAS.366.499} De Lucia G., Springel V., White S.~D.~M., Croton
D.~J, 2006, MNRAS, 366, 499

\bibitem[\protect\citeauthoryear{De Lucia
\& Blaizot}{2007}]{2007MNRAS.375....2D} De Lucia G., Blaizot J., 2007, MNRAS, 375, 2


\bibitem[\protect\citeauthoryear{De Lucia
\& Helmi}{2008}]{2008MNRAS.391...14D} De Lucia G., Helmi A., 2008, MNRAS, 391, 14

\bibitem[\protect\citeauthoryear{de Rossi, Tissera,
\& Scannapieco}{2007}]{2007MNRAS.374..323D} de Rossi M.~E., Tissera P.~B., Scannapieco C., 2007, MNRAS, 374, 323

\bibitem[\protect\citeauthoryear{Edvardsson et
al.}{1993}]{1993A&AS..102..603E} Edvardsson B., Andersen J., Gustafsson B., Lambert D.~L., Nissen P.~E., Tomkin J., 1993, A\&AS, 102, 603

\bibitem[\protect\citeauthoryear{Eggen, Lynden-Bell,
\& Sandage}{1962}]{1962ApJ...136..748E} Eggen O.~J., Lynden-Bell D., Sandage A.~R., 1962, ApJ, 136, 748

\bibitem[\protect\citeauthoryear{Erb et al.}{2006}]{2006ApJ...644..813E} 
Erb D.~K., Shapley A.~E., Pettini M., Steidel C.~C., Reddy N.~A., 
Adelberger K.~L., 2006, ApJ, 644, 813 

\bibitem[\protect\citeauthoryear{Erb et al.}{2006}]{2006ApJ...646..107E}
Erb D.~K., Steidel C.~C., Shapley A.~E., Pettini M., Reddy N.~A.,
Adelberger K.~L., 2006, ApJ, 646, 107

\bibitem[\protect\citeauthoryear{Finlator 
\& Dav{\'e}}{2008}]{2008MNRAS.385.2181F} Finlator K., Dav{\'e} R., 2008, MNRAS, 385, 2181

\bibitem[\protect\citeauthoryear{Font et 
al.}{2006}]{2006ApJ...638..585F} Font A. S., Johnston K. V., Bullock J. S.,
Robertson B. E., 2006, ApJ, 638, 585

\bibitem[\protect\citeauthoryear{Gao et al.}{2004}]{2004MNRAS.355..819G}
Gao L., White S.~D.~M., Jenkins A., Stoehr F., Springel V., 2004, MNRAS,
355, 819

\bibitem[\protect\citeauthoryear{Guesten
\& Mezger}{1982}]{1982VA.....26..159G} Guesten R., Mezger P.~G., 1982, VA, 26, 159

\bibitem[\protect\citeauthoryear{Hammer et al.}{2007}]
  {2007ApJ...662..322H} Hammer F., Puech M., Chemin L., Flores H., Lehnert
  M. D., 2007, ApJ, 662, 322 

\bibitem[\protect\citeauthoryear{Ivezic et al.}{2008}]{2008arXiv0804.3850I}
Ivezic Z., et al., 2008, ApJ, 684, 2871

\bibitem[\protect\citeauthoryear{Just 
\& Jahreiss}{2007}]{2007arXiv0706.3850J} Just A., Jahreiss H., 2007, arXiv, arXiv:0706.3850 

\bibitem[\protect\citeauthoryear{Kazantzidis et
al.}{2008}]{2008ApJ...688..254K} Kazantzidis S., Bullock J.~S., Zentner
A.~R., Kravtsov A.~V., Moustakas L.~A., 2008, ApJ, 688, 254


\bibitem[\protect\citeauthoryear{Kulkarni
\& Heiles}{1987}]{1987ASSL..134...87K} Kulkarni S.~R., Heiles C., 1987, ASSL, 134, 87

\bibitem[\protect\citeauthoryear{Lee et al.}{2006}]{2006ApJ...647..970L} 
Lee H., Skillman E.~D., Cannon J.~M., Jackson D.~C., Gehrz R.~D., Polomski 
E.~F., Woodward C.~E., 2006, ApJ, 647, 970 

\bibitem[\protect\citeauthoryear{Lequeux et 
al.}{1979}]{1979A&A....80..155L} Lequeux J., Peimbert M., Rayo J.~F., Serrano A., Torres-Peimbert S., 1979, A\&A, 80, 155 

\bibitem[\protect\citeauthoryear{Li
\& White}{2008}]{2008MNRAS.384.1459L} Li Y.-S., White S.~D.~M., 2008, MNRAS, 384, 1459



\bibitem[\protect\citeauthoryear{Lineweaver}{1999}]{1999Sci...284.1503L}
Lineweaver C.~H., 1999, Sci, 284, 1503

\bibitem[\protect\citeauthoryear{Maller}{2005}]{2005pgqa.conf..237M} Maller 
A.~H., 2005, pgqa.conf, 237

\bibitem[\protect\citeauthoryear{Moore 
\& Davis}{1994}]{1994MNRAS.270..209M} Moore B., Davis M., 1994, MNRAS, 270, 209

\bibitem[\protect\citeauthoryear{Navarro}{2004}]{2004ASSL..319..655N}
Navarro J.~F., 2004, ASSL, 319, 655

\bibitem[\protect\citeauthoryear{Pedersen et 
al.}{2006}]{2006NewA...11..465P} Pedersen K., Rasmussen J., Sommer-Larsen 
J., Toft S., Benson A.~J., Bower R.~G., 2006, NewA, 11, 465 

\bibitem[\protect\citeauthoryear{Perryman et
al.}{2001}]{2001A&A...369..339P} Perryman M.~A.~C., et al., 2001, A\&A, 369, 339

\bibitem[\protect\citeauthoryear{Savaglio et 
al.}{2005}]{2005ApJ...635..260S} Savaglio S., et al., 2005, ApJ, 635, 260

\bibitem[\protect\citeauthoryear{Searle
\& Zinn}{1978}]{1978ApJ...225..357S} Searle L., Zinn R., 1978, ApJ, 225, 357

\bibitem[\protect\citeauthoryear{Sch{\"o}del et
al.}{2002}]{2002Natur.419..694S} Sch{\"o}del R., et al., 2002, Natur, 419,
694

\bibitem[\protect\citeauthoryear{Simien
\& de Vaucouleurs}{1986}]{1986ApJ...302..564S} Simien F., de Vaucouleurs G., 1986, ApJ, 302, 564

\bibitem[\protect\citeauthoryear{Spergel et
al.}{2003}]{2003ApJS..148..175S} Spergel D.~N., et al., 2003, ApJS, 148,
175

\bibitem[\protect\citeauthoryear{Springel et
al.}{2005}]{2005Natur.435..629S} Springel V., et al., 2005, Natur, 435, 629

\bibitem[\protect\citeauthoryear{Steinmetz et
al.}{2006}]{2006AJ....132.1645S} Steinmetz M., et al., 2006, AJ, 132, 1645

\bibitem[\protect\citeauthoryear{Tissera, De Rossi, 
\& Scannapieco}{2005}]{2005MNRAS.364L..38T} Tissera P.~B., De Rossi M.~E., Scannapieco C., 2005, MNRAS, 364, L38 

\bibitem[\protect\citeauthoryear{Tremonti et
al.}{2004}]{2004ApJ...613..898T} Tremonti C.~A., et al., 2004, ApJ, 613,
898

\bibitem[\protect\citeauthoryear{Vivas \& Zinn}{2006}]
{2006AJ...132..714V} Vivas A.~K., Zinn R., 2006, AJ, 132,
714

\bibitem[\protect\citeauthoryear{Wang et al.}{2001}]{2001ApJ...555L..99W} 
Wang Q.~D., Immler S., Walterbos R., Lauroesch J.~T., Breitschwerdt D., 
2001, ApJ, 555, L99

\end{thebibliography}
\end{document}